\documentclass[preprint,showpacs,preprintnumbers,amsmath,amssymb]{revtex4}

\pdfoutput=1
\usepackage{graphicx}% Include figure files
\usepackage{dcolumn}% Align table columns on decimal point
\usepackage{bm}% bold math
\usepackage{amssymb}

\begin{document}
\title{
Effects of applied fields on quantum coupled double-well systems
%with Razavy's potential 
}% Force line breaks with \\
%\footnote{arXiv:1108:1156}

\author{Hideo Hasegawa}
\altaffiliation{hideohasegawa@goo.jp}
\affiliation{Department of Physics, Tokyo Gakugei University,  
Koganei, Tokyo 184-8501, Japan}%

\date{\today}% It is always \today, today,

%\keywords{nonequilibrium statistics,
%spatial correlation}%Use showkeys class option if keyword
\begin{abstract}
Effects of time-dependent applied fields on
quantum coupled double-well (DW) systems with Razavy's hyperbolic potential
have been studied.
By solving the Schr\"{o}dinger equation for the DW system, 
we have obtained time-dependent
occupation probabilities of the eigenstates, from which expectation values
of positions $x_1$ and $x_2$ of particles ($\langle x_1+x_2 \rangle$),
the correlation ($\Gamma(t)$) and the concurrence ($C(t)$) 
expressing a degree of the entanglement of the coupled DW system, are obtained.
Analytical expressions for $\langle x_1+x_2 \rangle$, $\Gamma(t)$ and $C(t)$
are derived with the use of the rotating-wave approximation (RWA) for sinusoidal fields.
Model calculations have indicated that 
$\langle x_1+x_2 \rangle$, $\Gamma(t)$ and $C(t)$ show very complicated time dependences.
Results of the RWA are in good agreement with exact ones evaluated by numerical methods
for cases of weak couplings and small applied fields in the near-resonant condition.
Applications of our method to step fields are also studied.

\vspace{0.5cm}

\end{abstract}
%03.65.-w  Quantum mechanics
%03.65.Ud  Entanglement and quantum nonlocality
%03.67.Bg  Entanglement production and manipulation
%03.67.Mn  Entanglement measures, witnesses, and other characterizations
%05.30.-d  Quantum statistical mechanics
%05.70.-a  thermodynamics                           
%05.40.-a  Fluctuation phenomenon, random process, noise
%05.10.Gg  Stochastic model
%05.45.-a  Nonlinear dynamics and chaos                          
%89.70.Cf  entropy and other measures of information

\maketitle
\newpage
\section{Introduction}

Extensive studies have been made for quantum double-well (DW) systems in physics 
and chemistry where a tunneling is one of intrigue quantum phenomena \cite{Tannor07}. 
Effects of applied fields on DW systems have been studied (for review see \cite{Grifoni98}). 
Various phenomena such as a coherent destruction of tunneling by applied fields were pointed
out \cite{Grossmann91}.
The two-level (TL) system which is a simplified model of a DW system, 
has been employed for a study on qubits which play important roles 
in quantum information and quantum computation. % \cite{Ref1}.
Many theoretical studies on effects of fields applied to single and coupled qubits 
have been reported with the use of the TL model \cite{Storcz01,Satanin12,Bina14,Pal14}.
In contrast to the simplified TL model, studies 
on coupled DW systems which are commonly described by the quartic potentials
are scanty \cite{Gupta06}, 
because a calculation of such a system is much tedious than 
that of the coupled TL model, even for the absence of applied fields. 
One of difficulties in studying coupled DW systems is that 
one cannot obtain exact eigenvalues and eigenfunctions 
of the Schr\"{o}dinger equation for quartic DW potential.
Then one has to apply various approximate approaches
such as perturbation and spectral methods to quartic DW models. 
Razavy \cite{Razavy80} proposed quasi-exactly solvable hyperbolic DW potential
for which one may exactly determine a part of whole eigenvalues and eigenfunctions.
A family of quasi-exactly solvable potentials has been investigated
\cite{Finkel99,Bagchi03}.

Recently the present author \cite{Hasegawa15} has investigated 
the relation between the entanglement and
the speed of evolution in coupled DW system described by Razavy's potential.
It would be interesting to study effects of applied fields on coupled DW systems 
with Razavy's potential, which is the purpose of the present study.
Some sophisticated methods like the Floquet approach have been developed 
in solving Schr\"{o}dinger equation for time-dependent periodic fields.
In order to treat the periodic as well as non-periodic dynamical fields,
we solve in this study the time-dependent Schr\"{o}dinger equation 
by a straightforward method.
An advantage of our approach is that we may exactly determine eigenvalues 
and eigenfunctions of driven coupled DW systems. 
We calculate expectation values of various quantities such as positions of
particles, the correlation and the concurrence which expresses a measure of
the entanglement of a coupled DW system.
Effects of applied fields are analytically studied with the use of the
rotating-wave approximation (RWA) which has been widely adopted for
sinusoidal periodic field, in particular for the TL model.
The validity of the RWA may be examined by a comparison between results of the RWA 
and exact ones evaluated by numerical methods.

The paper is organized as follows.
In Sec. II, the calculation method employed in our study is explained with
a brief review on Razavy's hyperbolic potential \cite{Razavy80}.
Equations of motion for populations of four energy levels
are obtained from the time-dependent Schr\"{o}dinger equation of
driven coupled DW systems. 
Expressions for expectation values of particle positions,
the correlation and the concurrence are calculated.
For sinusoidal fields, we present their analytical expressions by using the RWA.
In Sec. III, we report model calculations with the use of the RWA and 
numerical methods when the sinusoidal fields are applied to the initial ground state.
In Sec. IV, calculations are made for sinusoidal fields applied 
to the initially wavepacket state.
Our method is applied also to the case of applied step fields.
Sec. V is devoted to our conclusion.

%\newpage

\section{Coupled double-well system with Razavy's potential}
\subsection{Calculation method}
We consider coupled two DW systems whose Hamiltonian is given by 
\begin{eqnarray}
H &=& H_0 + H_C+ H_I,
\label{eq:A1}
\end{eqnarray}
where
\begin{eqnarray}
H_0 &=& \sum_{n=1}^2 \left[ -\frac{\hbar^2}{2m} \frac{\partial^2}{\partial x_n^2}
+ V(x_n )\right], 
\label{eq:A2}\\%
H_C &=& - g x_1 x_2, 
\label{eq:A3}\\
H_I &=& -(x_1+x_2) F(t), 
\label{eq:A4}\\
V(x) &=& \frac{\hbar^2}{2m}
\left[\frac{\xi^2}{8} \:{\rm cosh} \:4x - 4 \xi \:{\rm cosh} \:2x- \frac{\xi^2}{8}
\right].
\label{eq:A5}
\end{eqnarray}
Here $x_1$ and $x_2$ stand for coordinates of two distinguishable particles of mass $m$,
$H_0$ signifies a DW system with Razavy's potential $V(x)$ \cite{Razavy80}, 
$H_C$ means the coupling term with an interaction $g$, and $H_I$ includes
the time-dependent applied field $F(t)$ 
whose explicit form will be given shortly [Eq. (\ref{eq:C1}) or (\ref{eq:H2})].
The case of $H_I$ which is more general than Eq. (\ref{eq:A4}) will be studied in the Appendix.
The potential $V(x)$ with $\hbar=m=\xi=1.0$ adopted in this study
is plotted in Fig. 1(a).
Minima of $V(x)$ locate at $x_s=\pm 1.38433$ with $V(x_s)=-8.125$
and its maximum is $V(x)=-2.0$ at $x=0.0$.

Firstly we consider only $H_0$ in Eq. (\ref{eq:A2}), whose eigenvalues are given \cite{Razavy80}
\begin{eqnarray}
\epsilon_0 &=& \frac{\hbar^2}{2m}\left[ -\xi -5 -2 \sqrt{4-2 \xi+\xi^2} \right], \\
\epsilon_1 &=& \frac{\hbar^2}{2m}\left[ \xi-5 -2 \sqrt{4+2 \xi+\xi^2} \right], \\
\epsilon_2 &=& \frac{\hbar^2}{2m}\left[ -\xi-5 +2 \sqrt{4-2 \xi+\xi^2} \right], \\
\epsilon_3 &=& \frac{\hbar^2}{2m}\left[ \xi-5 +2 \sqrt{4+2 \xi+\xi^2} \right], 
\end{eqnarray}
and whose eigenfunctions are given by 
\begin{eqnarray}
\phi_0(x) &=& A_0 \; e^{-\xi \:{\rm cosh} \:2x/4} \left[3 \xi \:{\rm cosh} \:x
+(4-\xi+2 \sqrt{4-2 \xi+\xi^2})\: {\rm cosh}\: 3x \right], \\
\phi_1(x) &=&  A_1 \;e^{-\xi \:{\rm cosh} \:2x/4} \left[3 \xi \:{\rm sinh}\: x
+(4+\xi+2 \sqrt{4+2 \xi+\xi^2})\: {\rm sinh} \:3x \right], \\
\phi_2(x) &=& A_2 \; e^{-\xi \:{\rm cosh} \:2x/4} \left[3 \xi \:{\rm cosh} \:x
+(4-\xi-2 \sqrt{4-2 \xi+\xi^2})\: {\rm cosh}\: 3x \right], \\
\phi_3(x) &=&  A_3 \;e^{-\xi \:{\rm cosh} \:2x/4} \left[3 \xi \:{\rm sinh}\: x
+(4+\xi-2 \sqrt{4+2 \xi+\xi^2})\: {\rm sinh} \:3x \right], 
\end{eqnarray}
$A_{n}$ ($n=0,1$) denoting normalization factors.
Eigenvalues for the adopted parameters are $\epsilon_0=-4.73205$, $\epsilon_1=-4.64575$,
$\epsilon_2=-1.26795$ and  $\epsilon_3=0.645751$.
Both $\epsilon_0$ and $\epsilon_1$ locate below $V(0)$ as shown by dashed curves in Fig. 1(a),
and $\epsilon_2$ and $\epsilon_3$ are far above $\epsilon_1$. In this study,
we take into account only the lowest two states with $\epsilon_0$ and $\epsilon_1$,
which is justified because of
$\epsilon_1-\epsilon_0$ ($=0.0863$) $\ll \epsilon_2-\epsilon_1$ ($=3.3778$).
Figure 1(b) shows eigenfunctions of $\phi_0(x)$ and $\phi_1(x)$, which 
are symmetric and anti-symmetric, respectively, with respect to the origin.

\begin{figure}
\begin{center}
\includegraphics[keepaspectratio=true,width=120mm]{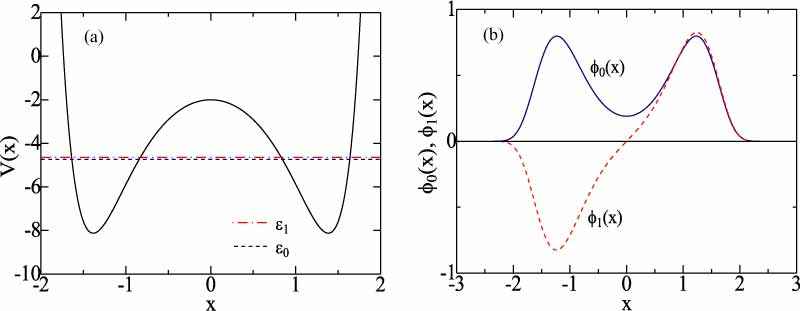}
\end{center}
\caption{
(Color online) 
(a) Razavy's DW potential $V(x)$ (solid curve),
dashed and chain curves expressing eigenvalues of $\epsilon_0$ and $\epsilon_1$,
respectively, for $\hbar=m=\xi=1.0$ [Eq.(\ref{eq:A5})].
(b) Eigenfunctions of $\phi_0(x)$ (solid curve) and $\phi_1(x)$ (dashed curve).
}
\label{fig1}
\end{figure}

Secondly we include the coupling term $H_C$ in Eq. (\ref{eq:A3}).
With basis states of $\phi_0(x_1)\phi_0(x_2)$,  $\phi_0(x_1)\phi_1(x_2)$,
$\phi_1(x_1)\phi_0(x_2)$ and  $\phi_1(x_1)\phi_1(x_2)$,
the energy matrix of the Hamiltonian of $H_0+H_C$ is expressed by
\begin{eqnarray}
{\cal H}_0 + {\cal H}_C &=& \left( {\begin{array}{*{20}c}
   {2 \epsilon_0 } & {0 } & {0 } & {-g \gamma^2} \\
   {0 } & {\epsilon_0 + \epsilon_1 } & {-g \gamma^2 } & {0} \\
   {0 } & {- g \gamma^2 } & {\epsilon_0 + \epsilon_1 } & {0} \\
   {- g \gamma^2 } & {0 } & {0 } & {2 \epsilon_1} \\   
\end{array}} \right),
\label{eq:A6}
\end{eqnarray}
with
\begin{eqnarray}
\gamma &=& \int_{-\infty}^{\infty} \phi_0(x)\: x \: \phi_1(x)\:dx=1.13823.
\label{eq:A7}
\end{eqnarray}
Eigenvalues of the energy matrix of ${\cal H}_0 + {\cal H}_C$ are given by
\begin{eqnarray}
E_0 &=& \epsilon -\sqrt{\delta^2+ g^2 \gamma^4}, 
\label{eq:A8}\\
E_1 &=& \epsilon - g \gamma^2, \\
E_2 &=& \epsilon + g \gamma^2, \\
E_3 &=& \epsilon + \sqrt{\delta^2+^2 \gamma^4},
\label{eq:A9}
\end{eqnarray}
where
\begin{eqnarray}
\epsilon &=& \epsilon_1+\epsilon_0=-9.3778, \\
\delta &=& \epsilon_1-\epsilon_0=0.0863.
\end{eqnarray}
Corresponding eigenfunctions are given by
%\begin{eqnarray}
%\Phi_0 &=& \cos \theta \:\vert 0,0 \rangle+ \sin \theta \: \vert 1,1 \rangle, \\
%\Phi_1 &=& \frac{1}{\sqrt{2}} \left( \vert 0,1 \rangle+ \vert 1,0 \rangle \right), \\
%\Phi_2 &=& \frac{1}{\sqrt{2}} \left(- \vert 0,1 \rangle+ \vert 1,0 \rangle \right), \\
%\Phi_3 &=& -\sin \theta \:\vert 0,0 \rangle+ \cos \theta \: \vert 1,1 \rangle,
%\end{eqnarray}
\begin{eqnarray}
\Phi_0(x_1,x_2) &=& \cos \theta \:\phi_0(x_1) \phi_0(x_2) 
+ \sin \theta \:\phi_1(x_1) \phi_1(x_2), 
\label{eq:A10a}\\
\Phi_1(x_1,x_2) &=& \frac{1}{\sqrt{2}} \left[ \phi_0(x_1) \phi_1(x_2)
+ \phi_1(x_1) \phi_0(x_2) \right], \\
\Phi_2(x_1,x_2) &=& \frac{1}{\sqrt{2}} \left[- \phi_0(x_1) \phi_1(x_2)
+ \phi_1(x_1) \phi_0(x_2) \right], \\
\Phi_3(x_1,x_2) &=& -\sin \theta \:\phi_0(x_1) \phi_0(x_2)
+ \cos \theta \: \phi_1(x_1) \phi_1(x_2),
\label{eq:A10b}
\end{eqnarray}
where
\begin{eqnarray}
%\tan \theta &=& \frac{g \gamma^2}
%{\delta + \sqrt{\delta^2+ g^2 \gamma^4} }, \\
\tan \:2 \theta &=& \frac{g \gamma^2}{\delta}.
\;\;\;\;\mbox{$\left(-\frac{\pi}{4} \leq \theta \leq \frac{\pi}{4} \right)$}
\label{eq:H7c}
\end{eqnarray}
We hereafter assume $g \geq 0$ \cite{Hasegawa15}.
The $g$ dependence of $E_{\nu}$ ($\nu=0-3$) is shown in Fig. 2 of Ref. \cite{Hasegawa15}.

Thirdly we take into account $H_I$ for an applied field in Eq. (\ref{eq:A4}).
The energy matrix of the time-dependent total Hamiltonian $H$ ($=H_0+H_C+H_I$) 
with basis states of $\phi_0(x_1)\phi_0(x_2)$,  $\phi_0(x_1)\phi_1(x_2)$,
$\phi_1(x_1)\phi_0(x_2)$ and  $\phi_1(x_1)\phi_1(x_2)$ is expressed by
\begin{eqnarray}
{\cal H}_0+{\cal H}_C+{\cal H}_I &=& \left( {\begin{array}{*{20}c}
   {2 \epsilon_0 } & {- \gamma F(t)} & {- \gamma F(t)} & {-g \gamma^2} \\
   {- \gamma F(t) } & {\epsilon_0 + \epsilon_1 } & {-g \gamma^2 } & {- \gamma F(t)} \\
   {- \gamma F(t)} & {- g \gamma^2 } & {\epsilon_0 + \epsilon_1 } & {- \gamma F(t)} \\
   {- g \gamma^2 } & {- \gamma F(t)} & {- \gamma F(t)} & {2 \epsilon_1} \\   
\end{array}} \right).
\label{eq:A11}
\end{eqnarray}

Alternatively, the energy matrix of $H$ may be expressed
with basis states of $\Phi_0$, $\Phi_1$, $\Phi_2$ and $\Phi_3$ 
in Eqs. (\ref{eq:A10a})-(\ref{eq:A10b}) by
\begin{eqnarray}
{\cal H} &=& \left( {\begin{array}{*{20}c}
   {E_0 } & {-\alpha F(t)} & {0} & {0} \\
   {-\alpha F(t) } & {E_1 } & {0} & {-\beta F(t)} \\
   {0 } & {0 } & {\;\;E_2\;\;} & {0} \\
   {0} & {-\beta F(t)} & {0 } & {E_3} \\   
\end{array}} \right), \nonumber \\
\label{eq:A12}
\end{eqnarray}
where
\begin{eqnarray}
\alpha &=& \sqrt{2} \gamma (\cos \theta + \sin \theta), 
\label{eq:A13} \\
\beta &=& \sqrt{2} \gamma (\cos \theta - \sin \theta).
\label{eq:A14}
\end{eqnarray}
In our following analysis, we adopt the energy matrix given by Eq. (\ref{eq:A12})
because it has more transparent physical meaning than Eq. (\ref{eq:A11}).
We expand the eigenstate $\Psi(x_1,x_2,t)$ of $H$ in terms of $\Phi_{\nu}(x_1,x_2)$ ($\nu=0-3$) 
with the time-dependent expansion coefficients $a_{\nu}(t)$ as
\begin{eqnarray}
\Psi(t) &=& \Psi(x_1,x_2,t)
= \sum_{\nu=0}^{3}\:a_{\nu}(t) \: \Phi_{\nu}(x_1,x_2) \:e^{-i E_{\nu} t/\hbar},
\label{eq:B1}
\end{eqnarray}
where expansion coefficients satisfy the relation
\begin{eqnarray}
\sum_{\nu=0}^3 \vert a_{\nu}(t) \vert^2 &=& 1.
\end{eqnarray}
The Schr\"{o}dinger equation: $i \hbar \:d\Psi(t)/dt=H \psi(t)$ becomes
\begin{eqnarray}
i \hbar \sum_{\nu=0}^3 \frac{d }{dt} 
\left[ a_{\nu}(t) \Phi_{\nu} \:e^{-i E_{\nu} t/\hbar} \right]
= \sum_{\nu=0}^3 a_{\nu}(t)E_{\nu} \Phi_{\nu} \:e^{-i E_{\nu}t/\hbar}
+ \sum_{\nu=0}^3 a_{\nu}(t) H_I \Phi_{\nu} \:e^{-i E_{\nu} t/\hbar}.
\label{eq:B2}
\end{eqnarray}
Multiplying $\Phi^*_{\mu}$ $(\mu=0-3)$ from the left side of Eq. (\ref{eq:B2}) 
and integrating it over $x_1$ and $x_2$, we obtain equations of motion for $a_{\mu}(t)$
\begin{eqnarray}
i \hbar \:\frac{d a_{\mu}(t)}{dt} 
&=& \sum_{\nu=0}^3 \langle \Phi_{\mu} \vert H_I \vert \Phi_{\nu}\rangle 
\:e^{-i \Delta_{\nu \mu}t} \:a_{\nu}(t), \hspace{1cm} \mbox{ $(\mu=0-3)$}
\label{eq:B3}
\end{eqnarray}
where
\begin{eqnarray}
\Delta_{\nu \mu} &=& \frac{E_{\nu}-E_{\mu}}{\hbar}.
\label{eq:B4}
\end{eqnarray}

With the use of the energy matrix in Eq. (\ref{eq:A12}), equations of motion for $a_{\mu}(t)$ become
[the argument $t$ in $a_{\nu}(t)$ is hereafter suppressed]
\begin{eqnarray}
i \hbar \:\frac{d a_0}{dt} &=& - \alpha F(t) \:e^{-i \Delta_{10} t} a_1, 
\label{eq:B5} \\
i \hbar \:\frac{d a_1}{dt} &=& -  \alpha F(t) \:e^{i \Delta_{10} t} a_0
- \beta F(t) e^{-i \Delta_{31}t}a_3 , 
\label{eq:B6} \\
i \hbar \:\frac{d a_2}{dt} &=& 0, 
\label{eq:B7} \\
i \hbar \:\frac{d a_3}{dt} &=& - \beta F(t) \:e^{i \Delta_{31} t} a_1.
\label{eq:B8}
\end{eqnarray}
%where
%\begin{eqnarray}
%\Delta_{\nu \mu} &=& \frac{E_{\nu}-E_{\mu}}{\hbar}.
%\end{eqnarray}

When we apply the sinusoidal field given by
\begin{eqnarray}
F(t) &=& f \:\sin \:\omega t,
\label{eq:C1}
\end{eqnarray}
Eqs. (\ref{eq:B5})-(\ref{eq:B8}) become
\begin{eqnarray}
\frac{d a_0}{dt} &=& \left( \frac{\alpha f}{2 \hbar} \right)
\left[ e^{i (\omega-\Delta_{10})t}- e^{-i (\omega+\Delta_{10})t} \right] \:a_1, 
\label{eq:C2}\\
\frac{d a_1}{dt} &=& \left( \frac{\alpha f}{2 \hbar} \right)
\left[ e^{-(\omega+\Delta_{10})t}- e^{-i (\omega-\Delta_{10})t} \right] \:a_0
+ \left( \frac{\beta f}{2 \hbar} \right)
\left[ e^{i (\omega-\Delta_{31})t}- e^{-i (\omega+\Delta_{31})t} \right] \:a_3, 
\label{eq:C3}\\
\frac{d a_2}{dt} &=& 0, 
\label{eq:C4}\\
\frac{d a_3}{dt} &=& \left( \frac{\beta f}{2 \hbar} \right)
\left[ e^{i (\omega+\Delta_{31})t}- e^{-i (\omega-\Delta_{31})t} \right] \:a_1, 
\label{eq:C5}
\end{eqnarray}
where $f$ and $\omega$ denote magnitude and frequency, respectively, of
the applied field.

\vspace{0.5cm}
\noindent
{\it Rotating-wave approximation (RWA)}

In the rotating-wave approximation (RWA) where
only terms with $(\omega - \Delta_{10})$ are taken into account 
in Eqs. (\ref{eq:C2})-(\ref{eq:C5}), we obtain 
\begin{eqnarray}
\frac{d a_0}{dt} &=& \left( \frac{ \alpha f}{2 \hbar} \right)
\:e^{i (\omega-\Delta_{10})t} \:a_1, 
\label{eq:D1} \\
\frac{d a_1}{dt} &=& - \left( \frac{ \alpha f}{2 \hbar} \right)
\:e^{-i (\omega-\Delta_{10})t} \:a_0,
\label{eq:D2} \\
\frac{d a_2}{dt} &=& \frac{d a_3}{dt}=0.
\label{eq:D3}
\end{eqnarray}
%The second-order DE for $a_1$ derived from Eqs. () and () becomes
%\begin{eqnarray}
%\frac{d^2 a_1}{dt^2} + i(\omega-\Delta_{10}) \:\frac{d a_1}{dt}
%+\left( \frac{\alpha f}{2 \hbar} \right)^2 a_1=0,
%\end{eqnarray}
For a given initial condition of $a_{\nu}(t)=a_{\nu}(0)$ at $t=0$ ($\nu=0-3$),
we obtain the solution of Eqs. (\ref{eq:D1})-(\ref{eq:D3})
\begin{eqnarray}
a_0(t) &=& \left( \frac{i \hbar}{\alpha f} \right)\: e^{i(\omega-\Delta_{10})t/2}
\left[A (\omega-\Delta_{10}-\Omega) \:e^{i \Omega t/2}
+B (\omega-\Delta_{10}+\Omega) \:e^{-i \Omega t/2} \right],
\label{eq:D4} \\
a_1(t) &=& e^{-i (\omega-\Delta_{10})t/2} 
\:(A \:e^{i \Omega t/2}+B \:e^{-i \Omega t/2}),
\label{eq:D5} \\
a_2(t) &=& a_2(0),
\label{eq:D6} \\
a_3(t) &=& a_3(0),
\label{eq:D7}
\end{eqnarray}
with 
\begin{eqnarray}
A &=& \left( \frac{i \alpha f}{2 \hbar \Omega} \right) a_0(0) 
+ \left( \frac{1}{2 \Omega} \right) (\omega-\Delta_{10}+\Omega) a_1(0),
\label{eq:D8} \\
B &=& -\left( \frac{i \alpha f}{2 \hbar \Omega} \right) a_0(0) 
- \left( \frac{1}{2 \Omega} \right) (\omega-\Delta_{10}-\Omega) a_1(0),
\label{eq:D9}
\end{eqnarray}
where $\Omega$ stands for Rabi's frequency given by
\begin{eqnarray}
\Omega &=& \sqrt{(\omega-\Delta_{10})^2 +(\alpha f/\hbar)^2}.
\label{eq:D10}
\end{eqnarray}
For a later purpose, we may rewrite $a_0(t)$ and $a_1(t)$ as 
\begin{eqnarray}
a_0(t) &=& r_0\:e^{i(\omega-\Delta_{10}-\Omega)t/2}
+s_0\:e^{i(\omega-\Delta_{10}+\Omega)t/2},
\label{eq:D11} \\
a_1(t) &=& r_1\:e^{-i(\omega-\Delta_{10}-\Omega)t/2}
+s_1\:e^{-i(\omega-\Delta_{10}+\Omega)t/2}, 
\label{eq:D12}
\end{eqnarray}
with
\begin{eqnarray}
r_0 &=& B \left( \frac{i \hbar}{\alpha f} \right) (\omega-\Delta_{10}+\Omega),
\label{eq:D13} \\
s_0 &=& A \left( \frac{i \hbar}{\alpha f} \right) (\omega-\Delta_{10}-\Omega),
\label{eq:D14} \\
r_1 &=& A, \;\;\; s_1=B,
\label{eq:D15}
\end{eqnarray}
where $r_{0}$, $s_0$, $r_1$ and $s_1$ are time independent.

\subsection{Various physical quantities}
Once time-dependent $a_{\nu}(t)$ are obtained from Eqs. (\ref{eq:B5})-(\ref{eq:B8}) 
or from Eqs. (\ref{eq:D11}) and (\ref{eq:D12}),
we may evaluate various physical quantities such as expectation values,
the correlation and concurrence.

\noindent
{\it (1) Expectation values}

Time-dependent expectation values of $\langle x_1 \rangle$
and $\langle x_2 \rangle$ are expressed by
\begin{eqnarray}
\langle x_1 \rangle &=& \int_{-\infty}^{\infty} \int_{-\infty}^{\infty} 
\Psi^*(x_1,x_2,t) x_1 \Psi(x_1,x_2,t) \:dx_1\:dx_2, \\ 
&=& \sqrt{2} \gamma\: [
(\cos \theta+\sin \theta) {\rm Re} \{a_0^* a_1 e^{-i\Delta_{10}t} -a_2^* a_3 e^{-i\Delta_{32}t}\}
\nonumber \\
&+&(\cos \theta-\sin \theta) {\rm Re} \{a_0^* a_2 e^{-i\Delta_{20}t} +a_1^* a_3 e^{-i\Delta_{31}t} \} ],
\label{eq:E1} \\
\langle x_2 \rangle &=& \sqrt{2} \gamma\: [
(\cos \theta+\sin \theta) {\rm Re} \{a_0^* a_1 e^{-i\Delta_{10}t} +a_2^* a_3 e^{-i\Delta_{32}t}\}
\nonumber \\
&-&(\cos \theta-\sin \theta) {\rm Re} \{a_0^* a_2 e^{-i\Delta_{20}t} +a_1^* a_3 e^{-i\Delta_{31}t} \} ],
\label{eq:E2} \\
\langle x_1 + x_2 \rangle &=&
2 \sqrt{2} \:\gamma (\cos \theta+\sin \theta)\:{\rm Re }\{a_0^* a_1e^{-i\Delta_{10}t} \}. 
\label{eq:E3} %
%\langle p_x + p_y \rangle &=& 
%2 \sqrt{2} \:\eta (\cos \theta+\sin \theta)\:{\rm Im }\{a_0^* a_1e^{-i\Delta_{10}t} \}, ???
\end{eqnarray}
%where
%\begin{eqnarray}
%\eta &=& \int_{-\infty}^{\infty} \phi_0(x) \;\frac{d \phi_1(x)}{dx} \:dx=0.09823.
%\label{eq:E4}
%\end{eqnarray}

Substituting Eqs. (\ref{eq:D11}) and (\ref{eq:D12}) to Eq. (\ref{eq:E3}), 
the expectation value in the RWA with $a_2(0)=a_3(0)=0$ is given by
\begin{eqnarray}
\langle x_1 \rangle_{RWA} &=& \langle x_2 \rangle_{RWA}
= \sqrt{2} \gamma (\cos \theta+\sin \theta)
\:{\rm Re}\{(r_0^* s_1+s_0^* r_1)\:e^{-i \omega t} \nonumber \\
&+& r_0^*r_1 \:e^{-i(\omega-\Omega)t}
+ s_0^* s_1 \:e^{-i(\omega+\Omega)t} \},
\label{eq:E5}
\end{eqnarray}
which includes time-dependent components with frequencies of $\omega \pm \Omega$ besides $\omega$ 
of the applied field.

\noindent
{\it (2) Correlation}

The correlation $\Gamma(t)$ is defined by \cite{Hasegawa15}
\begin{eqnarray}
\Gamma(t)^2 &=& \vert \int_{-\infty}^{\infty} \int_{-\infty}^{\infty} 
\Psi^*(x_1,x_2, 0) \:\Psi(x_1,x_2,t)\;dx_1\:dx_2 \: \vert^2, \\
&=& \vert \; \; a_0^*(0) a_0(t) + \sum_{\nu=1}^3 \:  a_{\nu}^*(0) a_{\nu}(t) 
\:e^{- i \Delta_{\nu 0} t} \: \vert^2,
\label{eq:F1}
\end{eqnarray}
which is unity at $t=0$.

In the RWA, the correlation is given by
\begin{eqnarray}
\Gamma_{RWA}(t)^2 &=& \vert a_0(0)\vert^2 \vert a_0(t)\vert^2+ \vert a_1(0)\vert^2 \vert a_1(t) \vert^2
+ 2 {\rm Re} \{ a_0(0) a_0^*(t) a_1^*(0)a_1(t) \:e^{-i \Delta_{10} t} \}.
\label{eq:F2}
\end{eqnarray}
With the use of Eqs. (\ref{eq:D11}) and (\ref{eq:D12}), the correlation in the RWA 
is expressed by
\begin{eqnarray}
\Gamma_{RWA}(t)^2 &=& (r_0^*+s_0^*)(r_0+s_0)( \vert r_0 \vert^2+\vert s_0 \vert^2
+ 2 {\rm Re}\{ s_0^* r_0 e^{-i \Omega t}\} ) \nonumber \\
&+& (r_1^*+s_1^*)(r_1+s_1)( \vert r_1 \vert^2+\vert s_1 \vert^2 
+ 2 {\rm Re}\{ r_1^* s_1 e^{-i \Omega t}\} ) \nonumber \\
&+& 2 {\rm Re} \{ (r_0+s_0)(r_1^*+s_1^*)[(s_0^* r_1+r_0^* s_1)\:e^{-i \omega t} \nonumber \\
&+& r_0^* r_1 \:e^{-i(\omega-\Omega)t}
+ s_0^* s_1\:e^{-i(\omega+\Omega)t}] \},
\label{eq:F3}
\end{eqnarray}
which consists of components with frequencies of $\omega$, $\Omega$ and $\omega \pm \Omega$.
When we take the average of $\Gamma_{RWA}(t)^2$ over a long period, oscillating
terms vanish and its average becomes
\begin{eqnarray}
\langle \Gamma_{RWA}(t)^2 \rangle_{av} &=& {\rm lim}_{T \rightarrow \infty}
\int_0^T \Gamma_{RWA}(t)^2\:dt, \\
&=&(r_0^*+s_0^*)(r_0+s_0)
( \vert r_0 \vert^2+\vert s_0 \vert^2)
+(r_1^*+s_1^*)(r_1+s_1)( \vert r_1 \vert^2+\vert s_1 \vert^2 ).
\label{eq:F4}
\end{eqnarray}

\noindent
{\it (3) Concurrence}

Substituting Eqs. (\ref{eq:A10a})-(\ref{eq:A10b}) into Eq. (\ref{eq:B1}), 
we obtain
\begin{eqnarray}
\vert \Psi \rangle &=& c_{00} \vert 0\;0 \rangle+ c_{01} \vert 0\;1 \rangle
+ c_{10} \vert 1\;0 \rangle + c_{11} \vert 1\;1 \rangle,
\label{eq:G1}
\end{eqnarray}
with
\begin{eqnarray}
c_{00} &=& a_0 \:\cos \theta \:e^{- i E_0 t}-a_3 \:\sin \theta \: e^{-i E_3 t}, 
\label{eq:G2a} \\
c_{01} &=&  \frac{1}{\sqrt{2}}(a_1 \:e^{- i E_1 t}-a_2 \:e^{- i E_2 t}), 
\label{eq:G2b} \\
c_{10} &=& \frac{1}{\sqrt{2}}(a_1\:e^{- i E_1 t}+a_2\:e^{- i E_2 t}),
\label{eq:G2c} \\
c_{11} &=& a_0\:\sin \theta\:e^{- i E_0 t}+a_3 \:\cos \theta\:e^{- i E_3 t},
\label{eq:G2d}
\end{eqnarray}
where $\vert k \;\ell \rangle=\phi_k(x_1) \phi_{\ell}(x_2)$ with $k, \ell=0,1$.
The concurrence $C$ of the state $\vert \Psi \rangle$ given by Eq. (\ref{eq:G1}) is defined by
\cite{Wootters01}
\begin{eqnarray}
C^2 &=& 4 \:\vert c_{00} c_{11}- c_{01} c_{10} \vert^2. %(\geq 0).
\label{eq:G3}
\end{eqnarray}
The state given by Eq. (\ref{eq:G1}) becomes factorizable if and only if the
relation: $c_{00} c_{11}- c_{01} c_{10} =0$ holds.
Substituting Eqs. (\ref{eq:G2a})-(\ref{eq:G2d}) into Eq. (\ref{eq:G3}), 
we obtain the concurrence \cite{Hasegawa15}
\begin{eqnarray}
C(t)^2 &=& \vert [a_0(t)^2 - a_3(t)^2 \:e^{-2i \Delta_{30} t}] \sin 2 \theta
+ 2 a_0(t) a_3(t) \:\cos 2\theta \:e^{-i \Delta_{30} t}
- a_1(t)^2 \:e^{-2i \Delta_{10} t} \nonumber \\
&+& a_2(t)^2 \:e^{-2 i \Delta_{20} t} \vert^2.
\label{eq:G4}
\end{eqnarray}

In the RWA with $a_2(0)=a_3(0)=0$, the concurrence is given by
\begin{eqnarray}
C_{RWA}(t)^2 &=& \vert a_0(t)^2 \sin 2 \theta
- a_1(t)^2 \:e^{-2i \Delta_{10} t} \vert^2.
\label{eq:G5}
\end{eqnarray}
With the use of Eqs. (\ref{eq:D11}) and (\ref{eq:D12}), Eq. (\ref{eq:G5}) becomes
\begin{eqnarray}
C_{RWA}(t)^2 &=& D_0-D_1 \sin 2\theta+ D_2 \sin^2 2 \theta,
\label{eq:G6}
\end{eqnarray}
with
\begin{eqnarray}
D_0 &=&\vert r_1 \vert^4 + \vert s_1 \vert^4 
+ 4 \vert r_1 \vert^2 \vert s_1 \vert^2 
+ 2 {\rm Re}\{ 2(\vert r_1 \vert^2+\vert s_1 \vert^2)r_1^* s_1\:e^{-i\Omega t}
+ r_1^{*2} s_1^2 \:e^{-2i\Omega t} \}, 
\label{eq:G7} \\
D_1 &=& 2 {\rm Re}\{(r_0^{*2} s_1^2+s_0^{*2}r_1^2+4 r_0^* s_0^* r_1 s_1)\:e^{-2i \omega t}
+ 2(r_0^* s_0^* r_1^2+r_0^{*2} r_1 s_1)\:e^{-i(2 \omega-\Omega)t}\nonumber \\
&+& 2(r_0^* s_0^* s_1^2+s_0^{*2} r_1 s_1)\:e^{-i(2 \omega +\Omega)t} 
+ r_0^{*2}r_1^2\:e^{-2i(\omega-\Omega)t}+s_0^{*2} s_1^2 \:e^{-2i(\omega+\Omega)t} \},
\label{eq:G8} \\
D_2 &=& \vert r_0 \vert^4 + \vert s_0 \vert^4 
+ 4 \vert r_0 \vert^2 \vert s_0 \vert^2 
+ 2 {\rm Re}\{ 2(\vert r_0 \vert^2+\vert s_0 \vert^2)s_0^* r_0\:e^{-i\Omega t}
+ s_0^{*2} r_0^2 \:e^{-2i\Omega t} \}, 
\label{eq:G9}
\end{eqnarray}
which include contributions from multiple components with frequencies of $\omega$, $2\omega$,
$\Omega$, $2 \Omega$, $\omega \pm \Omega$, $2\omega \pm \Omega$ and $2(\omega \pm \Omega)$.
The concurrence averaged over a long period is given by
\begin{eqnarray}
\langle C_{RWA}(t)^2 \rangle_{av} &=& (\vert r_1 \vert^4 + \vert s_1 \vert^4 
+ 4 \vert r_1 \vert^2 \vert s_1 \vert^2)
+ \sin^2 2 \theta \:(\vert r_0 \vert^4 + \vert s_0 \vert^4 
+ 4 \vert r_0 \vert^2 \vert s_0 \vert^2 ).
\label{eq:G10}
\end{eqnarray}

\begin{figure}
\begin{center}
\includegraphics[keepaspectratio=true,width=120mm]{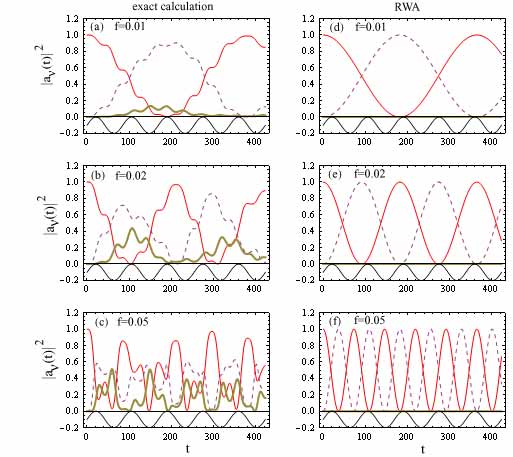}
\end{center}
\caption{
(Color online) 
Time developments of $\vert a_0(t) \vert^2$ (solid curves), $\vert a_1(t) \vert^2$ (dashed curves) 
and $\vert a_3(t) \vert^2$ (bold solid curve) % $F(t)=f\: \sin \omega t$
for (a) $f=0.01$, (b) $f=0.02$ and (c) $f=0.05$ in exact calculations,
and those for (d) $f=0.01$, (e) $f=0.02$ and (f) $f=0.05$ in the RWA
($g=0.01$ and $\omega=\Delta_{10}$), bottom curves in (a)-(f) expressing applied fields.
}
\label{figA}
\end{figure}

\begin{figure}
\begin{center}
\includegraphics[keepaspectratio=true,width=160mm]{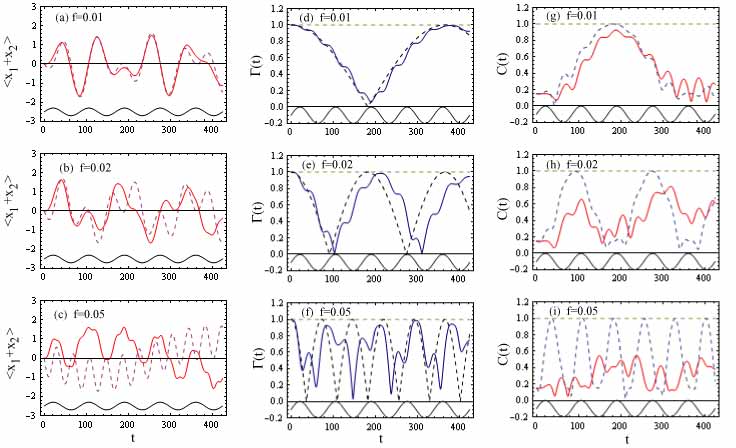}
\end{center}
\caption{
(Color online) 
(a)-(c) $\langle x_1+x_2 \rangle$ for (a) $f=0.01$, (b) $f=0.02$ and (c) $f=0.05$;
(d)-(f) $\Gamma(t)$ for (d) $f=0.01$, (e) $f=0.02$ and (f) $f=0.05$;
(g)-(i) $C(t)$ for (g) $f=0.01$, (h) $f=0.02$ and (i) $f=0.05$,
solid and dashed curves expressing results of exact and RWA calculations, respectively
($g=0.01$ and $\omega=\Delta_{10}$).
Bottom curves in (a)-(i) denote applied fields.
}
\label{figB}
\end{figure}

\section{Model calculations}

Assuming the initial ground state given by
\begin{eqnarray}
a_0(0)&=& 1,\;\;a_1(0)= a_2(0)=a_3(0)=0,
\label{eq:H1} 
\end{eqnarray}
we have made numerical calculations by changing model parameters of $f$, $g$ and $\omega$.

\begin{figure}
\begin{center}
\includegraphics[keepaspectratio=true,width=120mm]{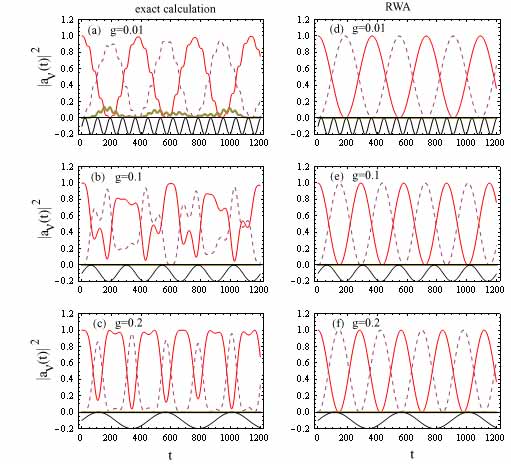}
\end{center}
\caption{
(Color online) 
Time developments of $\vert a_0(t) \vert^2$ (solid curves), $\vert a_1(t) \vert^2$ (dashed curves) 
and $\vert a_3(t) \vert^2$ (bold solid curve)
for (a) $g=0.01$, (b) $g=0.1$, and (c) $g=0.2$ in exact calculations,
and those for (d) $g=0.01$, (e) $g=0.1$, and (f) $g=0.2$ in the RWA
($f=0.01$ and $\omega=\Delta_{10}$),
bottom curves in (a)-(f) expressing applied fields.
}
\label{figC}
\end{figure}

\subsection{$f$ dependence}
Figures \ref{figA}(a), (b) and (c) show time developments of populations 
of $\vert a_{\nu}(t) \vert^2$ in levels $\nu$ ($=0, 1, 3$) for $f=0.0$, 0.01 and 0.02, 
respectively with $g=0.01$ and $\omega=\Delta_{10}$ (=0.07431) obtained
by numerically solving Eqs. (\ref{eq:C2})-(\ref{eq:C5})
which is hereafter referred to as an exact calculation:
note that $a_{2}(t)=0$ because of $d \:a_2(t)/dt=0$ and $a_2(0)=0$.
For comparison, relevant results obtained in the RWA are plotted in Figs. \ref{figA}(d)-(f).
Exact calculations in Fig. \ref{figA}(a) show that for a field with $f=0.01$, 
magnitude of $\vert a_0(t) \vert^2$ is decreased
while that of $\vert a_1(t) \vert^2$ is increased at $t \simeq 0$ with 
small $\vert a_3(t) \vert^2$, which is similar to results in the RWA shown in Fig. 2(d).
For $f=0.02$, however, magnitude of $ \vert a_3(t) \vert^2$ in exact calculations
becomes appreciable [Fig. \ref{figA}(b)] whereas it is vanishing in the RWA [Fig. \ref{figA}(e)].
A comparison between Figs. \ref{figA}(c) and (f) show that
the difference between an exact calculation and the RWA is evident for $f=0.05$.

Figures \ref{figB}(a), (b) and (c) express time dependences of $\langle x_1+x_2 \rangle$
for $f=0.01$, 0.02 and 0.05, respectively, obtained by exact calculations (solid curves)
and the RWA (dashed curves). 
We note that $\langle x_1+x_2 \rangle$ shows a complicated time dependence which arises 
from a superposition of multiple motions with frequencies of $\omega$ and $\omega \pm \Omega$
as the RWA analysis shows.
This analysis may be applied to case of $f=0.01$ and $f=0.02$. For a larger
$f=0.05$, however, $\langle x_1+x_2 \rangle$ of the RWA is very different from that of
an exact calculation [Fig. \ref{figB}(c)].

Figures \ref{figB}(d), (e) and (f) show the correlation $\Gamma(t)$ calculated for 
$f=0.01$, 0.02 and 0.05, respectively, with $g=0.01$.
RWA analysis shows that $\Gamma(t)^2$ oscillates with 
frequencies of $\omega$, $\Omega$ and $\omega \pm \Omega$.

The concurrence $C(t)$ calculated for $f=0.01$, 0.02 and 0.05 are shown 
Figs. \ref{figB}(g), (h) and (i), respectively.
At $t=0$, $C(0)=0.1485$. 
$C(t)$ shows complicated time dependence because $C(t)^2$ includes
superposed oscillations with frequencies of $\omega$, $2\omega$,
$\Omega$, $2 \Omega$, $\omega \pm \Omega$, $2\omega \pm \Omega$ and $2(\omega \pm \Omega)$,
as analyzed by the RWA which is expected to be valid for a small $f$.

\subsection{$g$ dependence}
Next we report calculated results when the coupling $g$ is varied.
Figure \ref{figC}(a), (b) and (c) show time developments of $\vert a_{\nu} \vert^2$
for $g=0.01$, $0.1$ and $0.2$, respectively, obtained by exact calculations
with $f=0.01$ and $\omega=\Delta_{10}$ (=0.07431, 0.02611 and 0.0140 for $g=0.01$, 0.1 and 0.2, respectively).
For comparison, relevant results obtained in the RWA are plotted in Figs. \ref{figC}(d)-(f).
A result of the RWA for $g=0.01$ in Fig. \ref{figC}(d) is in fairly good agreement with that of
an exact calculation in Fig. \ref{figC}(a).
For $g=0.1$, however, an agreement between exact and RWA calculations is not satisfactory
[Fig. \ref{figC}(b)].
For $g=0.2$ result of the RWA is quite different from that of an exact calculation [Fig. \ref{figC}(c)].

\begin{figure}
\begin{center}
\includegraphics[keepaspectratio=true,width=160mm]{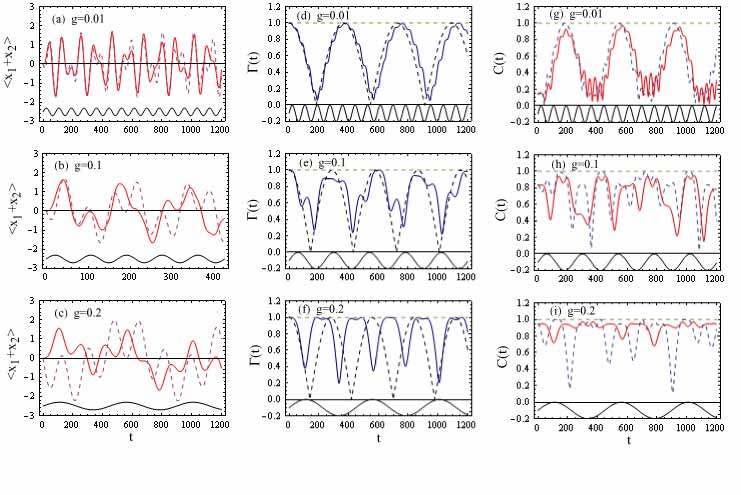}
\end{center}
\caption{
(Color online) 
(a)-(c) $\langle x_1+x_2 \rangle$ for (a) $g=0.01$, (b) $g=0.1$ and (c) $g=0.2$; 
(d)-(f) $\Gamma(t)$ for (d) $g=0.01$, (e) $g=0.1$, and (f) $g=0.2$;
(g)-(i) $C(t)$ for (g) $g=0.01$, (h) $g=0.1$, and (i) $g=0.2$,
solid and dashed curves expressing results of exact and RWA calculations, respectively
($f=0.01$ and $\omega=\Delta_{10}$).
Bottom curves in (a)-(i) denote applied fields.
}
\label{figD}
\end{figure}

The difference between results of $\vert a_{\nu} \vert^2$ in an exact calculation and
the RWA reflects on expectation values of $\langle x_1+x_2 \rangle$ shown 
in Figs.\ref{figD}(a)-(f).
We note that although expectation values in an exact calculation and the RWA are in fairly good agreement 
for $g=0.01$, both the results are rather different for $g=0.1$ and 0.2.

Figs. \ref{figD}(d)-(f) and Figs. \ref{figD}(g)-(i) show
the correlation and concurrence, respectively,
obtained by an exact calculation and the RWA for several $g$ values with $f=0.01$.
We note in Figs. \ref{figD}(g)-(i) that the concurrence becomes larger for larger $g$.
In fact, the time-averaged $\langle C_{RWA}(t)^2 \rangle_{av}$ given by Eq. (\ref{eq:G10}) is 
0.383265, 0.634747 and 0.712556 for $f=0.01$, 0.02 and 0.05, respectively.

\subsection{$\omega$ dependence}
\begin{figure}
\begin{center}
\includegraphics[keepaspectratio=true,width=120mm]{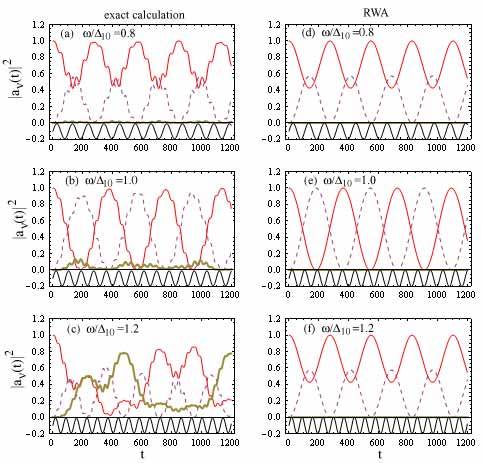}
\end{center}
\caption{
(Color online) 
Time developments of $\vert a_0(t) \vert^2$ (solid curves), $\vert a_1(t) \vert^2$ (dashed curves) 
and $\vert a_3(t) \vert^2$ (bold solid curve)
for (a) $\omega/\Delta_{10}=0.8$, (b) $1.0$, and (c) $1.2$ in exact calculations,
and those for (d) $\omega/\Delta_{10}=0.8$, (e) $1.0$, and (f) $1.2$ in the RWA
($f=g=0.01$),
bottom curves in (a)-(f) expressing applied fields.
}
\label{figE}
\end{figure}

Although we have so far assumed that $\omega$ is equal to $\Delta_{10}$, its value 
is changed in this subsection.
Figures \ref{figE}(a), (b) and (c) show exact calculations of 
time-dependent level populations $\vert a_{\nu}(t) \vert^2$ 
for $\omega/\Delta_{10}=0.8$, 1.0 and 1.2, respectively, with $f=g=0.01$.
For comparison, relevant results in the RWA are plotted in Figs. \ref{figE}(d)-(f).
In the resonant condition with $\omega = \Delta_{10}$, $\vert a_0(t) \vert^2$
and $\vert a_1(t) \vert^2$ oscillate between 0 and 1. It is not the case
in the off-resonant condition with $\omega \neq \Delta_{10}$.
We note in Figs. 10(d) and (f) that a result for $\omega/\Delta_{10}=0.8$
is the same as that for $\omega/\Delta_{10}=1.2$ in the RWA. This is because 
$\Omega$ is expressed in term of $(\omega-\Delta_{10})^2$ in Eq. (\ref{eq:D10}).
In exact calculations, however, a result for $\omega/\Delta_{10}=0.8$ is
different from that for $\omega/\Delta_{10}=1.2$ as shown in Figs. \ref{figE}(a) and (c).
In particular, an exact calculation for $\omega/\Delta_{10}=1.2$ in Fig. \ref{figE}(c)
shows a peculiar time dependence, which is quite different from a result of
the RWA in Fig. \ref{figE}(f).

\begin{figure}
\begin{center}
\includegraphics[keepaspectratio=true,width=160mm]{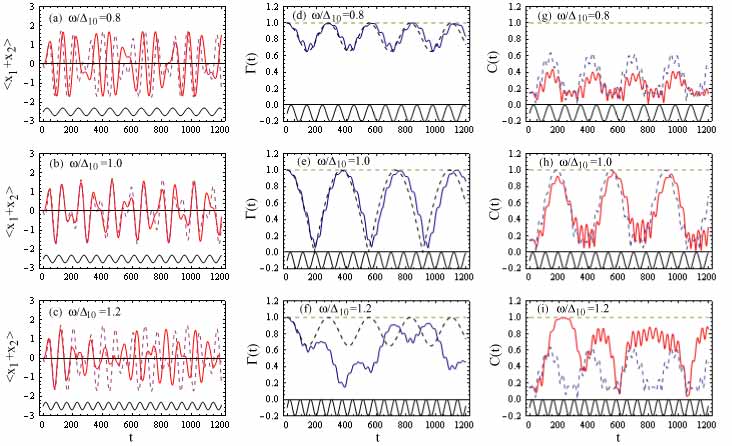}
\end{center}
\caption{
(Color online) 
(a)-(c) $\langle x_1+x_2 \rangle$ 
for (a) $\omega/\Delta_{10}=0.8$, (b) $1.0$, and (c) $1.2$;
(d)-(f) $\Gamma(t)$ for (a) $\omega/\Delta_{10}=0.8$, (b) $1.0$, and (c) $1.2$;
(g)-(i) $C(t)$ for (a) $\omega/\Delta_{10}=0.8$, (b) $1.0$, and (c) $1.2$,
solid and dashed curves expressing results of exact and RWA calculations, respectively
($f=0.01$ and $g=0.01$). 
Bottom curves in (a)-(i) denote applied fields.
}
\label{figF}
\end{figure}

Time dependences of $\langle x_1+x_2 \rangle$ for $\omega/\Delta_{10}=0.8$, 1.0 and 1.2
are shown in Fig. \ref{figF}(a), (b) and (c) obtained by exact calculations (solid curves)
and the RWA (dashed curves).

Figures \ref{figF}(d)-(f) and Figs. \ref{figF}(g)-(i) show
the correlation function and concurrence, respectively,
for $\omega/\Delta_{10}=0.8$, 1.0 and 1.2 calculated by exact calculations (solid curves)
and the RWA (dashed curves).
Again results of $\Gamma(t)$ and $C(t)$ in the RWA for $\omega/\Delta_{10}=0.8$ are the same as
those for $\omega/\Delta_{10}=1.2$.
Results of the RWA are not in good agreement with those of exact calculations
except for the resonant condition of $\omega = \Delta_{10}$ with a small $f$ and a weak $g$.

\begin{figure}
\begin{center}
\includegraphics[keepaspectratio=true,width=120mm]{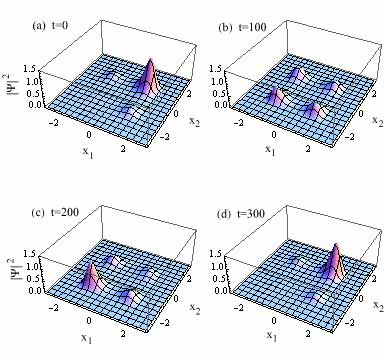}
\end{center}
\caption{
(Color online) 
Magnitudes of wavefunctions $\vert \Psi(x_1, x_2, t) \vert^2$ 
at (a) $t=0.0$, (b) $t=100.0$, (c) $t=200.0$ and (d) $t=300.0$
with $f=0.02$, $g=0.01$ and $\omega=\Delta_{10}$ for the initial wavepacket 
given by Eq. (\ref{eq:H3})
}
\label{figW}
\end{figure}

\section{Discussion}
\subsection{Initial wavepacket state}

It would be interesting to study effects of fields applied to
the initial wavepacket state given by
\begin{eqnarray}
a_0(0)&=& a_1(0)=\frac{1}{\sqrt{2}},\;\;a_2(0)=a_3(0)=0.
\label{eq:H3}
\end{eqnarray}
Figure \ref{figW}(a) shows the 3D plot of magnitude of $\vert \Psi(x_1, x_2, t) \vert^2$ at $t=0.0$
as functions of $x_1$ and $x_2$ with $f=0.02$, $g=0.01$ and $\omega=\Delta_{10}$, 
which initially has a main peak at $(x_1, x_2)=(1.23, 1.23)$. 
Figures \ref{figW}(b)-(d) will be explained shortly.

\begin{figure}
\begin{center}
\includegraphics[keepaspectratio=true,width=120mm]{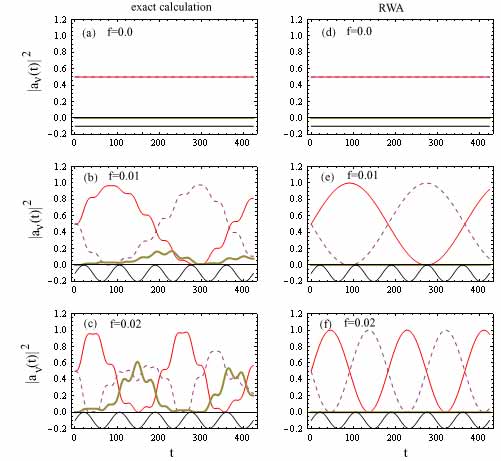}
\end{center}
\caption{
(Color online) 
Time developments of $\vert a_0(t) \vert^2$ (solid curves), $\vert a_1(t) \vert^2$ (dashed curves) 
and $\vert a_3(t) \vert^2$ (bold solid curve) % $F(t)=f\: \sin \omega t$
for (a) $f=0.0$, (b) $f=0.01$ and (c) $f=0.02$ in exact calculations,
and those for (d) $f=0.0$, (e) $f=0.01$ and (f) $f=0.02$ in the RWA; 
the initial wavepacket given by Eq. (\ref{eq:H3}) is adopted 
with $g=0.01$ and $\omega=\Delta_{10}$. Bottom curves in (a)-(f) express applied fields.
}
\label{figG}
\end{figure}

Figures \ref{figG}(a), (b) and (c) show time developments of populations 
of $\vert a_{\nu}(t) \vert^2$ in levels $\nu$ ($=0, 1, 3$) for $f=0.0$, 0.01 and 0.02, 
respectively with $g=0.01$ and $\omega=\Delta_{10}$ ($=0.07431$) obtained
by exact calculations.
Relevant results obtained in the RWA are plotted in Figs. \ref{figG}(d)-(f).
For $f=0.0$, the populations are time independent, 
$\vert a_{0}(t)\vert^2= \vert a_{1}(t) \vert^2=1/2$ 
and $\vert a_2(t) \vert^2 = \vert a_3(t) \vert^2 =0.0$.
For $f=0.01$, $\vert a_0(t) \vert^2$ is initially increased while $\vert a_0(t) \vert^2$ is
decreased from their initial values of $1/2$ by an applied field. 
A result of the RWA for $f=0.01$ in Fig. \ref{figG}(e) is in fairly good agreement with 
that of an exact calculation in Fig. \ref{figG}(b).
However, Figs. \ref{figG}(c) and (f) show that for $f=0.02$ both results are different, 
in particular, a population of $\vert a_3(t) \vert^2$ becomes appreciable in an exact calculation
while it is vanishing in the RWA.

3D plots of magnitudes of wavefunctions $\vert \Psi(x_1, x_2, t) \vert^2$ 
at $t=100.0$,  200.0 and 300.0 are presented in Figs. \ref{figW}(b), (c) and (d),
respectively.
At $t=100.0$, the wavepacket has four small peaks at $(x_1, x_2)=(\pm 1.23, \pm 1.23)$
and $(x_1, x_2)=(\pm 1.23, \mp 1.23)$ [Fig.\ref{figW}(b)]. 
At $t=200.0$, it has again four small peaks at the same positions as at $t=100.0$ 
[Fig.\ref{figW}(c)]
although a peak at $(x_1, x_2)=(-1.23, -1.23)$ is the highest among the four. 
The wavepacket returns approximately to the initial state at $t=300.0$ as shown 
in Fig. \ref{figW}(d), which is similar to Fig. \ref{figW}(a) at $t=0.0$. 

\begin{figure}
\begin{center}
\includegraphics[keepaspectratio=true,width=160mm]{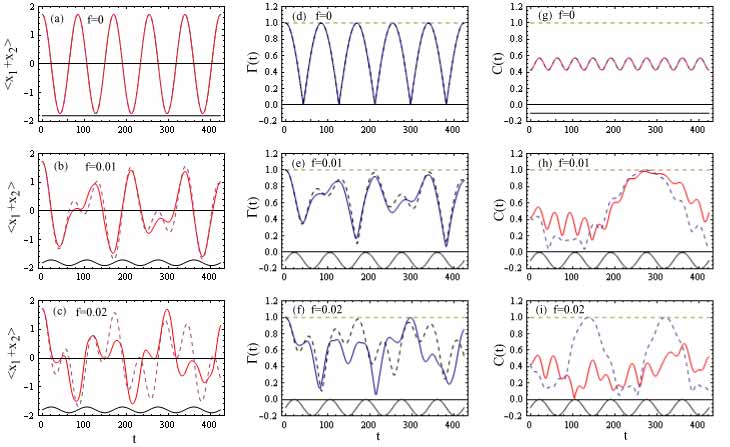}
\end{center}
\caption{
(Color online) 
(a)-(c) $\langle x_1+x_2 \rangle$ for (a) $f=0.0$, (b) $f=0.01$ and (c) $f=0.02$;
(d)-(f) $\Gamma(t)$ for (d) $f=0.0$, (e) $f=0.01$ and (f) $f=0.02$;
(g)-(i) $C(t)$ for (g) $f=0.0$, (h) $f=0.01$ and (i) $f=0.02$,
solid and dashed curves expressing results of exact and RWA calculations, respectively,
for the initial wavepacket given by Eq. (\ref{eq:H3}) ($g=0.01$ and $\omega=\Delta_{10}$). 
Bottom curves in (a)-(i) denote applied fields.
}
\label{figH}
\end{figure}

Figures \ref{figH}(a), (b) and (c) express time dependences of $\langle x_1+x_2 \rangle$
for $f=0.0$, 0.01 and 0.02, respectively, obtained by exact calculations (solid curves)
and the RWA (dashed curves). We note in Fig. \ref{figH}(a) that 
$\langle x_1+x_2 \rangle$ for $f=0.0$ shows the simple sinusoidal time dependence,
which express a tunneling of wavepacket between two minima of DW potential.
When a field with $f=0.01$ is applied, both results of $\langle x_1+x_2 \rangle$ 
obtained by exact calculations and the RWA show complicated time dependences 
[Fig. \ref{figH}(b)].
Although the both results are in good agreement for $f=0.0$ and 0.01, 
results of exact and the RWA calculations become different for a larger $f=0.02$ 
as shown in Figs. \ref{figH}(c).
We note that a tunneling becomes difficult by applied fields, which is nothing but
a suppression of tunneling by coherent fields \cite{Grossmann91}.

Figures \ref{figH}(d), (e) and (f) show the time dependence of correlation $\Gamma(t)$
for $f=0.0$, 0.01 and 0.02, respectively, with $g=0.01$ obtained by an exact calculation
(solid curves) and the RWA (dashed curves).
$\Gamma(t)$ for the adopted wavepacket with $f=0.0$ is given by \cite{Hasegawa15}
\begin{eqnarray}
\Gamma(t)^2&=& \frac{1}{2}[1+\cos (\Delta_{10}t)], 
\end{eqnarray}
which leads to a sinusoidal oscillation with a period of $2 \pi/\Delta_{10}=84.55$.
Figure \ref{figH}(e) shows that $\Gamma(t)$ for $f=0.01$ shows complicated
time dependences and that a result of the RWA is in fairly good agreement with that
of an exact calculation. For a larger $f=0.02$, however, a result of the RWA becomes different 
from that of the exact calculation [Fig. \ref{figH}(f)].

Figures \ref{figH}(g), (h) and (i) show the concurrence $C(t)$
for $f=0.0$, 0.01 and 0.02, respectively, with $g=0.01$ obtained by an exact calculation
(solid curves) and the RWA (dashed curves).
$C(t)$ of the adopted wave packet for $f=0.0$ is given by \cite{Hasegawa15}
\begin{eqnarray}
C(t)^2 &=& \frac{1}{4}[1+ \sin^2 (2 \theta) -2 \sin (2 \theta) \cos (2 \Delta_{10} t)],
\end{eqnarray}
which yields $C(0)=0.42577$ at $t=0.0$ for $g=0.01$.
$C(t)$ for $f=0.0$ oscillates with the period of $\pi/\Delta_{10}=42.28$
as shown in Fig. \ref{figH}(g).
Figures \ref{figH}(h) and (i) show that
for finite $f=0.01$ and 0.02, $C(t)$ has complicated time dependence,
just as $\Gamma(t)$.

\subsection{Step field}
\begin{figure}
\begin{center}
\includegraphics[keepaspectratio=true,width=120mm]{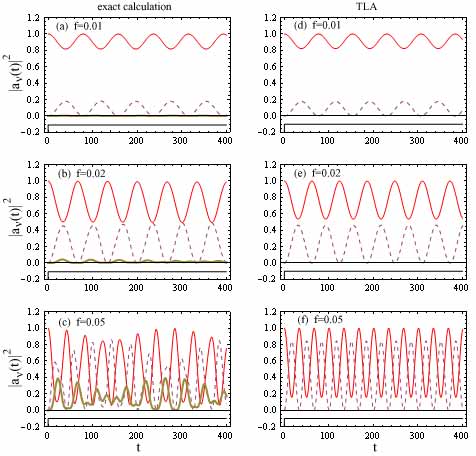}
\end{center}
\caption{
(Color online) 
Time developments of $\vert a_0(t) \vert^2$ (solid curves), 
$\vert a_1(t) \vert^2$ (dashed curves) 
and $\vert a_3(t) \vert^2$ (bold solid curve) % $F(t)=f\: \sin \omega t$
for the applied step field with (a) $f=0.01$, (b) $f=0.02$ and (c) $f=0.05$ 
in exact calculations,
and those for (d) $f=0.01$, (e) $f=0.02$ and (f) $f=0.05$ in the TLA ($g=0.01$),
bottom curves in (a)-(f) expressing applied step fields.
}
\label{figK}
\end{figure}

\begin{figure}
\begin{center}
\includegraphics[keepaspectratio=true,width=160mm]{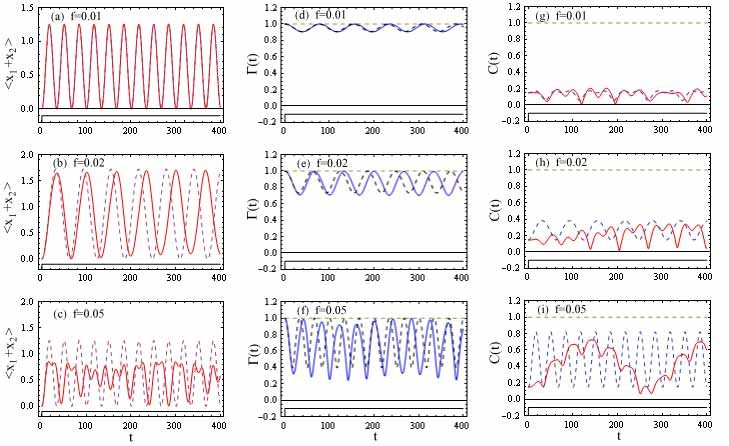}
\end{center}
\caption{
(Color online) 
(a)-(c) $\langle x_1+x_2 \rangle$ for (a) $f=0.01$, (b) $f=0.02$ and (c) $f=0.05$;
(d)-(f) $\Gamma(t)$ for (d) $f=0.01$, (e) $f=0.02$ and (f) $f=0.05$;
(g)-(i) $C(t)$ for (g) $f=0.01$, (h) $f=0.02$ and (i) $f=0.05$,
solid and dashed curves expressing results of exact and TLA calculations, respectively,
($g=0.01$). Bottom curves in (a)-(i) denote applied step fields.
}
\label{figJ}
\end{figure}
Although sinusoidal periodic fields have been so far adopted, we may apply our 
method to any time-dependent field. As a typical example, we employ a step field given by
\begin{eqnarray}
F(t) &=& f\: \Theta(t),
\label{eq:H2}
\end{eqnarray}
where $\Theta(t)$ denotes the Heaviside function.
%The initial condition is assumed to be given by Eq. (\ref{eq:H1}).

We first try to obtain approximate analytical expressions of $a_{\nu}(t)$ ($\nu=0-3$)
for an applied step field with the use of the two-level approximation (TLA) 
where contributions only from two terms of $a_0(t)$ and $a_1(t)$ are included.
Equations (\ref{eq:B5})-(\ref{eq:B8}) for $t > 0$ become
\begin{eqnarray}
i \hbar \:\frac{d a_0}{dt} &=& - \alpha f \:e^{-i \Delta_{10} t} a_1, 
\label{eq:J1} \\
i \hbar \:\frac{d a_1}{dt} &=& -  \alpha f \:e^{i \Delta_{10} t} a_0.
\label{eq:J2} \\
i \hbar \:\frac{d a_2}{dt} &=& 
i \hbar \:\frac{d a_3}{dt} = 0.
\label{eq:J2b}
\end{eqnarray}
Solutions of Eqs. (\ref{eq:J1})-(\ref{eq:J2b}) are given by
\begin{eqnarray}
a_0(t) &=& \left( \frac{\hbar}{2 \alpha f} \right)
[(\Delta_{10}-\Omega_s) A \:e^{-i(\Delta_{10}+\Omega_s)t/2}
+(\Delta_{10}+\Omega_s) B \:e^{-i(\Delta_{10}-\Omega_s)t/2} ],
\label{eq:J3} \\
a_1(t) &=& A \:e^{i(\Delta_{10}-\Omega_s)t/2}+B \:e^{i(\Delta_{10}+\Omega_s)t/2},
\label{eq:J4} \\
a_2(t) &=& a_2(0), \\
a_3(t) &=& a_3(0),
\end{eqnarray}
with
\begin{eqnarray}
\Omega_s &=& \sqrt{\Delta_{10}^2+ 4(\alpha f/\hbar)^2},
\label{eq:J5}
\end{eqnarray}
where $A$ and $B$ stand for integration constants.
For assumed initial conditions $a_0(0)=1$ and $a_1(0)=a_2(0)=a_3(0)=0$ which yield
$A=-B=-\alpha f/\hbar \Omega_s$, solutions are given by
\begin{eqnarray}
a_0(t) &=& \left( \frac{1}{2 \Omega_s} \right)
[(\Delta_{10}+\Omega_s) \:e^{-i(\Delta_{10}-\Omega_s)t/2}
-(\Delta_{10}-\Omega_s) \:e^{-i(\Delta_{10}+\Omega_s)t/2} ],
\label{eq:J6} \\
a_1(t) &=& -\left( \frac{\alpha f}{\hbar \Omega_s} \right)
[e^{i(\Delta_{10}-\Omega_s)t/2}- e^{i(\Delta_{10}+\Omega_s)t)/2}],
\label{eq:J7} \\
a_2(t) &=& a_3(t) =0.
\end{eqnarray}
Then $\vert a_0(t) \vert^2$ and $\vert a_1(t) \vert^2$ are expressed by superposed 
oscillations with frequencies of $\Omega_s \pm \Delta_{10}$.

Figures \ref{figK}(a), (b) and (c) show $\vert a_{\nu} \vert^2$ 
of exact calculations for step fields with $f=0.01$, 0.02 and 0.05,
respectively, which are applied to the initial ground state given by Eq. (\ref{eq:H1}); 
relevant results in the TLA are plotted in Figs. \ref{figK}(d), (e) and (f).
When a step field is applied, $\vert a_{\nu}(t) \vert^2$ begin to oscillate. 
Equations (\ref{eq:J6}) and (\ref{eq:J7}) may explain
time dependences of $\vert a_0(t) \vert^2$ and $\vert a_0(t) \vert^2$
for small $f=0.01$ and 0.02.
They are, however, not valid for $f=0.05$, for which $\vert a_{3}(t) \vert^2$ has
an appreciable magnitude while it is assumed to be zero in the TLA.

Time dependences of $\langle x_1 +x_2 \rangle$, $\Gamma(t)$ and $C(t)$
are plotted in Figs. \ref{figJ}(a)-(c), Figs. \ref{figJ}(d)-(f)
and Figs. \ref{figJ}(g)-(i), respectively, where solid and dashed curves
denote results in exact calculations and in the TLA, respectively.
Their time dependences for $f=0.01$ and 0.02 may be approximately elucidated in the TLA
given by Eqs. (\ref{eq:J6}) and (\ref{eq:J7}), although they
are not applicable to the case of $f=0.05$.

\section{Conclusion}
We have studied effects of applied fields in quantum coupled DW system
with the use of an exactly solvable  Razavy's potential \cite{Razavy80}. 
From the Schr\"{o}dinger equation for the driven DW system, we have obtained equations of motion 
for populations of the four levels. Model calculations of expectation values 
$\langle x_1+x_2 \rangle$, correlation $\Gamma(t)$ and concurrence $C(t)$
for applied sinusoidal fields show very complicated time dependence. 
Their time dependence may be analytically understood within the RWA in cases of 
a weak coupling and a small field in the near-resonance of
$\vert \omega-\Delta_{10} \vert \ll \omega$. Otherwise, results of the RWA 
are not in good agreement with exact numerical calculations.
It is indispensable to develop an analytical method going beyond the RWA
for the DW system, just as for the TL model \cite{Ashhab07,Werlang08,Son09,He12}.
In the present study, we do not take into account environmental effects which are expected
to play important roles in real DW systems. These are left as our future subjects.

\begin{acknowledgments}
This work is partly supported by a Grant-in-Aid for Scientific Research from 
Ministry of Education, Culture, Sports, Science and Technology of Japan.  
\end{acknowledgments}

%\vspace{0.5cm}
\appendix*

%\newpage
\section{A. General driving fields}
\renewcommand{\theequation}{A\arabic{equation}}
\setcounter{equation}{0}
We consider the input term which is more general than Eq. (\ref{eq:A4}), as given by
\begin{eqnarray}
H_I &=& -x_1 F_1(t) -x_2 F_2(t),
\label{eq:X1} 
\end{eqnarray}
$F_n(t)$ ($n=1, 2$) expressing a time-dependent field applied to the $n$th DW system.
The energy matrix of the time-dependent total Hamiltonian $H$ ($=H_0+H_C+H_I$) 
with basis states of $\Phi_0$, $\Phi_1$, $\Phi_2$ and $\Phi_3$ is given by
\begin{eqnarray}
{\cal H} &=& \left( {\begin{array}{*{20}c}
   {E_0 } & {-\frac{\alpha}{2} [F_1(t)+F_2(t)]} & {-\frac{\beta}{2} [F_1(t)-F_2(t)]} & {0} \\
   {-\frac{\alpha}{2} [F_1(t)+F_2(t)] } & {E_1 } & {0} & {-\frac{\beta}{2} [F_1(t)+F_2(t)]} \\
   {-\frac{\beta}{2} [F_1(t)-F_2(t)] } & {0 } & {\;\;E_2\;\;} & {\frac{\alpha}{2} [F_1(t)-F_2(t)]} \\
   {0} & {-\frac{\beta}{2} [F_1(t)+F_2(t)] } & {\frac{\alpha}{2} [F_1(t)-F_2(t)]} & {E_3} \\   
\end{array}} \right), \nonumber \\
\label{eq:X2}
\end{eqnarray}
where $\alpha$ and $\beta$ are given by Eqs. (\ref{eq:A13}) and (\ref{eq:A14}), respectively.
Note that in the case of $F_1(t)=F_2(t)=F(t)$, Eq. (\ref{eq:X2}) becomes Eq. (\ref{eq:A12}).

With the use of Eqs. (\ref{eq:B3}) and (\ref{eq:X2}), 
equations of motion for $a_{\mu}(t)$ ($\mu=0-3$) become
\begin{eqnarray}
i \hbar \:\frac{d a_0}{dt} &=& - \frac{\alpha}{2} [F_1(t)+F_2(t)] \:e^{-i \Delta_{10} t} a_1 
- \frac{\beta}{2} [F_1(t)-F_2(t)] \:e^{-i \Delta_{20} t} a_2, 
\label{eq:X5} \\
i \hbar \:\frac{d a_1}{dt} &=& - \frac{\alpha}{2} [F_1(t)+F_2(t)] \:e^{i \Delta_{10} t} a_0 
- \frac{\beta}{2} [F_1(t)+F_2(t)] \:e^{-i \Delta_{31} t} a_3, 
\label{eq:X6} \\
i \hbar \:\frac{d a_2}{dt} &=& - \frac{\beta}{2} [F_1(t)-F_2(t)] \:e^{i \Delta_{20} t} a_0 
+ \frac{\alpha}{2} [F_1(t)-F_2(t)] \:e^{-i \Delta_{32} t} a_3, 
\label{eq:X7} \\
i \hbar \:\frac{d a_3}{dt} &=& - \frac{\beta}{2} [F_1(t)+F_2(t)] \:e^{i \Delta_{31} t} a_1 
+ \frac{\alpha}{2} [F_1(t)-F_2(t)] \:e^{i \Delta_{32} t} a_2. 
\label{eq:X8}
\end{eqnarray}

\begin{figure}
\begin{center}
\includegraphics[keepaspectratio=true,width=60mm]{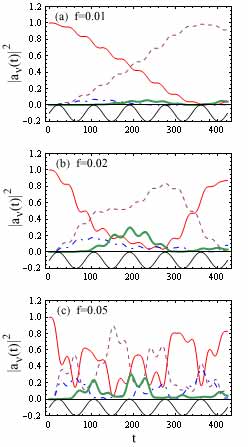}
\end{center}
\caption{
(Color online) 
Time developments of $\vert a_0(t) \vert^2$ (solid curves), 
$\vert a_1(t) \vert^2$ (dashed curves), $\vert a_2(t) \vert^2$ (chain curves)
and $\vert a_3(t) \vert^2$ (bold solid curve) % $F(t)=f\: \sin \omega t$
with (a) $f=0.01$, (b) $f=0.02$ and (c) $f=0.05$ 
in exact calculations,
bottom curves in (a)-(c) expressing applied fields of $F_1(t)$ [$F_2(t)=0$].
}
\label{figX}
\end{figure}

\begin{figure}
\begin{center}
\includegraphics[keepaspectratio=true,width=160mm]{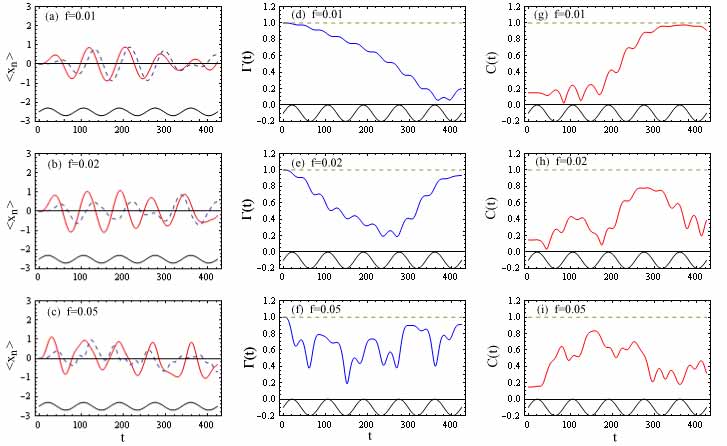}
\end{center}
\caption{
(Color online) 
(a)-(c): Time dependences of $\langle x_1 \rangle$ (solid curves) and $\langle x_2 \rangle$ (dashed curves)  
for (a) $f=0.01$, (b) $0.02$ and (c) $0.05$;
(d)-(f): $\Gamma(t)$ for (d) $f=0.01$, (e) $0.02$ and (f) $0.05$;
(g)-(i): $C(t)$ for (g) $f=0.01$, (h) $0.02$ and (i) $0.05$ ($g=0.01$).
Bottom curves in (a)-(i) denote applied fields of $F_1(t)$ [$F_2(t)=0$].
}
\label{figY}
\end{figure}

In the symmetric case of $F_1(t)=F_2(t)=F(t)$, Eqs. (\ref{eq:X5})-(\ref{eq:X8}) 
reduce to Eqs. (\ref{eq:B5})-(\ref{eq:B8}). 
In the anti-symmetric case of $F_1(t)=-F_2(t)=F(t)$, Eqs. (\ref{eq:X5})-(\ref{eq:X8}) become
\begin{eqnarray}
i \hbar \:\frac{d a_0}{dt} &=& - \beta F(t) \:e^{-i \Delta_{20} t} a_2, 
\label{eq:X9} \\
i \hbar \:\frac{d a_1}{dt} &=& 0, 
\label{eq:X10} \\
i \hbar \:\frac{d a_2}{dt} &=& - \beta F(t) \:e^{i \Delta_{20} t} a_0
+ \alpha F(t) e^{-i \Delta_{32}t}a_3 , 
\label{eq:X11} \\
i \hbar \:\frac{d a_3}{dt} &=& \alpha F(t) \:e^{i \Delta_{32} t} a_2.
\label{eq:X12}
\end{eqnarray}
for which we obtain $a_1(t)=a_1(0)$.

When fields are applied only to the first DW system with $F_1(t) \neq 0$ and $F_2(t)=0$, 
Eqs. (\ref{eq:X5})-(\ref{eq:X8}) yield
\begin{eqnarray}
i \hbar \:\frac{d a_0}{dt} &=& - \frac{\alpha F_1(t)}{2} \:e^{-i \Delta_{10} t} a_1 
- \frac{\beta F_1(t)}{2} \:e^{-i \Delta_{20} t} a_2, 
\label{eq:X13} \\
i \hbar \:\frac{d a_1}{dt} &=& - \frac{\alpha F_1(t)}{2} \:e^{i \Delta_{10} t} a_0 
- \frac{\beta F_1(t)}{2} \:e^{-i \Delta_{31} t} a_3, 
\label{eq:X14} \\
i \hbar \:\frac{d a_2}{dt} &=& - \frac{\beta F_1(t)}{2} \:e^{i \Delta_{20} t} a_0 
+ \frac{\alpha F_1(t)}{2}  \:e^{-i \Delta_{32} t} a_3, 
\label{eq:X15} \\
i \hbar \:\frac{d a_3}{dt} &=& - \frac{\beta F_1(t)}{2} \:e^{i \Delta_{31} t} a_1 
+ \frac{\alpha F_1(t)}{2} \:e^{i \Delta_{32} t} a_2.
\label{eq:X16}
\end{eqnarray}

For applied fields of $F_1(t)=f \:\sin \omega t$ and $F_2(t)=0$ with small $f$ and weak $g$ 
where $a_2(t)$ and $a_3(t)$ are negligible as will be shown shortly,
we may take into account only $a_0(t)$ and $a_1(t)$. Then
Eqs. (\ref{eq:X13})-(\ref{eq:X16}) within the RWA become
\begin{eqnarray}
\frac{d a_0}{dt} &=& \left( \frac{ \alpha f}{4 \hbar} \right)
\:e^{i (\omega-\Delta_{10})t} \:a_1, 
\label{eq:X17} \\
\frac{d a_1}{dt} &=& - \left( \frac{ \alpha f}{4 \hbar} \right)
\:e^{-i (\omega-\Delta_{10})t} \:a_0,
\label{eq:X18} \\
\frac{d a_2}{dt} &=& \frac{d a_3}{dt}=0,
\label{eq:X19}
\end{eqnarray}
which are equivalent to Eqs. (\ref{eq:D1})-(\ref{eq:D3}) for the symmetric driving fields
if we read $f/2 \rightarrow f$.

Figures \ref{figX}(a)-(c) show time developments of $\vert a_{\nu}(t) \vert^2$ ($\nu=0-3$)
calculated for $f=0.01$, 0.02 and 0.05, respectively, with $g=0.01$ and $\omega=\Delta_{10}$
when fields of $F_1(t)=f \sin \omega t$ and $F_2(t)=0$ are applied
to the system with an initial stable state given by Eq. (\ref{eq:H1}).
When a field with $f=0.01$ is applied, $\vert a_0(t) \vert^2$ is decreased from unity while
$\vert a_1(t) \vert^2$ begins to increase at $t \sim 0.0$ with
vanishingly small $\vert a_2(t) \vert^2$ and $\vert a_3(t) \vert^2$. 
For larger fields with $f=0.02$ and 0.05, however, magnitudes of $\vert a_2(t) \vert^2$ 
and $\vert a_3(t) \vert^2$ become appreciable.
Results in Fig. \ref{figX}(b) with $f=0.02$ are not dissimilar to those in Fig. \ref{figA}(a) 
for the symmetrical field with $f=0.01$, as mentioned above.

Time dependences of $\langle x_1 \rangle$ (solid curves) 
and $\langle x_2 \rangle$ (dashed curves) for $f=0.01$, 0.02 and 0.05 are plotted
in Figs. \ref{figY}(a), (b) and (c), respectively, where
$\langle x_1 \rangle$ is different from $\langle x_2 \rangle$:
the latter seems a little delayed than the former.
Note that $\langle x_2 \rangle = 0.0$ for $g=0.0$ for which the two DW systems are
decoupled (related results not shown).  

Relevant results of $\Gamma(t)$ for $f=0.01$, 0.02 and 0.05 are shown in
Figs. \ref{figY}(d), (e) and (f), respectively, while $C(t)$ are plotted
in Figs. \ref{figY}(g), (h) and (i). They show very complicated time dependence.
It is impossible to theoretically elucidate time dependences of $\vert a_{\nu}(t) \vert^2$,
$\langle x_n \rangle$ ($n=1, 2$), $\Gamma(t)$ and $C(t)$ except for cases with very small $f$ 
and weak $g$.

%\newpage

\end{document}